# Gravity is indeed the driving force for fault pattern formation and westward displacement of the lithosphere

Andrés R. R. Papa

*Coordenação de Geofísica, Observatório Nacional ON-MCT, Rua General José Cristino 77, São Cristovão, Rio de Janeiro, 20921-400 RJ, BRAZIL.*

*Department of Mathematics and Computation, State University of Rio de Janeiro, Estrada Resende –Riachuelo s/n, Morada da Colina, Resende, 27523-000 RJ, BRAZIL.*

**Gravity influence of the Sun and the Moon on the Earth is the cause of the fault pattern on Earth's surface we observe nowadays and also of the westward displacement of the lithosphere. A somewhat related hypotheses advanced in the past[1] have been that tidal torque of the Moon may be significant. However, it has been argued that it is untenable[2] essentially on the grounds that the viscosity of the Earth's mantle is far too high in comparison with the forces involved. To state this it was assumed that the Earth's relevant characteristic is that of two spherical shells separated by a viscous liquid, the external shell rotating at a constant angular velocity with respect to the inner shell. Here I show that this picture is not quite right and that a more careful look leads us to one in which both shells rotates at different mean angular velocities and with a forward-backward sequence of movements relative to each other. In this way what was supposed to be the principal factor against gravity influence becomes the main favourable one.**

Earthquakes are one of the most explored phenomena in Earth nature because of their intrinsic interest but fundamentally because of the threat that they represent for the human kind. However, the motor force of seismic activity is far from being clearly stated. The common factor to almost all the explanations[3] that have been forwarded as possible driving forces for plate tectonics appears to be the gravitation. Gravitation coming from the Earth is supposed to be the cause in the case of plates pushed from ridges, pulled from trenches and dragged from beneath by the mantle. The same is the case for plates sliding downhill from ridges. Earth gravity is still involved in some theories on the pressure driven phase transitions from basalt rocks (of approximate density $3g/cm^3$) to eclogite (density $3.6g/cm^3$) that would cause some plates to sink into others. Moon's gravity has been advanced as a possible force to drag plates from above. Here I introduce the possible gravitational effects, but now from the Sun (and also from the Moon), on the seismic activity on the Earth. The Sun gravitational force on a piece of ground is about 150 times the force that the Moon exerts on the same piece in the average.

Earth rotation defines a privileged direction on space for phenomena that take place on Earth: the rotation axis direction. The other special "direction" is the Earth surface: plates are forced to move in directions more or less "parallel" to this surface.

Suppose four neighbouring hypothetical plates on the equatorial section of the Earth as seen from the North Pole. Two of them centred on the line that joins the centre of the Earth and the centre of the Sun (or Moon) and two on the line perpendicular to the first that passes by the centre of the Earth. As plates are forced to move on the surface, each plate experiments periodically indirect forces coming from its two neighbours. They have sine-like time dependence. Note that this implies that near the Equator line, there should be a preferential orientation North-South for Earth surface defects as ridges and trenches owing to the hit-stretch sequence at which plates are submitted. This is the case[4], for example, in the central Atlantic ridges as well as in the American border of the great Pacific plate. An exception to this rule will happen when there are stronger forces involved in that regions as is the case, for example, of very rigid plates.

Suppose now two neighbouring hypothetical plates, one of them centred on the Equator and the other in one of the tropics. The surface component of the acceleration (force/mass ratio) is the same for both plates. However, the movement possibilities for regions on the Equator should be greater than, for example, regions near Tropics. This difference could cause sliding regions. Near the Equator then, there should be also expected a concomitant fault pattern parallel to the Equator, they should be preferentially sliding regions. This type of pattern is also observed[4] in the near-equatorial part of the central Atlantic and through the whole eastern part of the Pacific.

Closer to the poles both effects are weaker. To reinforce this idea let us consider a section of the Earth containing the two poles. If we artificially divide each pole in two halves it is not too difficult to realize that, independently of the chosen instant, the tangential force exerted by the Sun or the Moon on each of the halves is the same. The important fact from this point of view is that the forces on all the semi-plates are practically equal, and more important, they are all in phase. The rapid conclusion of this is that in Polar Regions there should be expected almost zero earthquake activity because of the lack of fault production. A polar plate is maintained as a single unit by its own position. The extension of the integral unbroken polar plates can be estimated to extend for about 25 degrees from each of the poles: the extreme points where the superficial component of the force does not differ in more than 10 percent of its maximum value. There is an amazing coincidence with this prediction in South Pole[4]. In North Pole[4] they are observed but for latitudes well beyond 25 degrees from the Pole indicating, probably, the existence of a huge very rigid plate. At the same time, both polar plates act as a hammer for sub-polar regions. In those regions there should be Earth surface defects of the type ridge or similar. In contrast with the equatorial regions now this type of defect should be observed in east-west directions, as is the case in practically all the extension[4] of the south regions of Pacific, Indian and Atlantic oceans. On the other hand the movement of polar plates on the sub-polar regions is similar to the roll of a wheel: at each instant the polar plate exerts a force on a narrow north-south strip. This "wheel-effect" will cause sliding regions oriented approximately in the north-south direction contrary to what is observed in equatorial regions (east-west). Such patterns are observed[4] in the western side of the South Pacific and, principally, along the Indian Ocean.

To close the analysis, consider two contiguous pieces A and B of spherical shells at different depths. By the same arguments as before, we can expect plate over plate sliding. This phenomenon could increase or decrease locally the effects of all that was explained above for plate movement on the surface. Let me analyse in more detail this situation, it will supply an extra support to the ideas presented up to now and bring the basis for the westward displacement of lithosphere. Suppose that pieces A and B belong to two spherical shells with effective elastic constants opposed to dislocations on each of the shells $k_1$ and $k_2$, respectively. It is not too difficult to show that under equal tangential force/mass (the case for gravity) ratios, the angular displacements, $\theta_1$ and $\theta_2$, of both pieces will be the same. This implies that there is not relative angular displacement between pieces A and B. But there is no reason to believe that $k_1=k_2$. On the contrary, there are physical and geometrical reasons that point to different elastic constants at different depths: diverse constitution of layers and more deformed layers when passing from a diameter $r$ to another $r´$, where $r´ > r$, as the closer the original layer to the centre of the sphere. From both reasons it should be expected $k_1 > k_2$. For a perfect sphere, without dissipation and for a long enough time, there is not a mean dislocation because all that happens at any instant will return to its original position when the pieces A and B pass over the opposite position on the sphere. Liquid and solid tidal budges supply the necessary asymmetry to do of this process an irreversible one.

The inclusion of a Newtonian fluid layer (asthenosphere) between the external shell (lithosphere) and the inner shell (mesosphere) does not disqualify the conclusions here advanced. The viscosity $\eta$ of the liquid shell will just, depending on its value, to reinforce or to weakening the related effects.

To illustrate this fact, but without pretending to drain the subject, I have simulated two pieces of spherical shells at different depth and submitted to periodical tangential forces (owing to the gravity of the Sun). They are also interconnected through a viscous force, with viscosity constant $\eta$. For both slabs the periodic force was set to expressions of the type $A(sin(\omega.t))$ where $A$ is the amplitude, $\omega$ the angular frequency and $t$ the time. To simulate the asymmetry introduced by tidal budges for the outer shell it was assumed a value $A=1$ for the positive parts of the sine function and to a fraction very close to 1 (I have simulated, for example, with values 0.999 and 0.99999, among others) in the negative portions. A similar procedure was adopted for the inner shell, but as it was not explicitly introduced the difference between the internal elastic forces on each shell (that would require a much more complicated simulation, for example, of a whole equatorial disc), this was artificially introduced by giving a value for the corresponding amplitude (positive portions of the sine function) slightly different from 1. The value of $A$ for the inner shell and for the negative portions of the sine function was determined by using the same proportion than for the outer shell.

The first result extracted from the simulations is a join angular velocity for both shells following some time dependence. This result should be expected (for example, in relation to the poles). However, the specific time dependence is not to be taken too seriously because the corresponding part of the simulation was not implemented in a detailed fashion. This is the reason to not further discuss this result here.

The second and essential result is that the shells realize a continuous forward-backward sequence of motions relative to each other with an average relative angular velocity different from zero. After many simulations with different parameters values it was clear that the average relative angular velocity $\dot{\theta}$ depends inversely on the viscosity coefficient $\eta$ ($k$ is a proportionality constant depending on the values of the other parameters in the simulations):

$$\dot{\theta} = \frac{k}{\eta} \qquad (1).$$

The above analysis implies that during a part of the day the inner shell helps the outer shell to rotate while during the other part opposes to its movement. There is a net dislocation between both shells because of the irreversibility of the process. On the other hand, Equation (1) assures the existence of this dislocation, without mattering the value of $\eta$. With relation to a fix coordinate system (*e.g.*, the poles where this effect is null) both shells rotate at different angular velocities in the same direction. Actually, more detailed simulations together with some known results (on hot spot data[2], for instance) can be used to determine viscosity values.

**Competing Interests statement**  The author declares that he has no competing financial interests.

**Correspondence** and requests for materials should be addressed to A.R.R.P. (e-mail: papa@on.br).